\documentstyle[12pt,aasms4]{article}
\received{}

\newcommand {\eg} {e.g.}		
\newcommand {\ie} {i.e.}		
\newcommand {\etal} {\it et al.}        

\begin{document}

\title{White Dwarfs in Globular Clusters: {\it HST} Observations of M4$\dagger$}

\author{Harvey B. Richer\altaffilmark{1}, Gregory G. Fahlman\altaffilmark{1}, Rodrigo A. Ibata\altaffilmark{1}, Carlton Pryor\altaffilmark{2}, Roger A. Bell\altaffilmark{3}, Michael Bolte\altaffilmark{4}, Howard E. Bond\altaffilmark{5}, William E. Harris\altaffilmark{6}, James E. Hesser\altaffilmark{7}, Steve Holland\altaffilmark{1}, Nicholas Ivanans\altaffilmark{1}, Georgi Mandushev\altaffilmark{1}, Peter B. Stetson\altaffilmark{7} \& Matt A. Wood\altaffilmark{8}}

\altaffiltext{1}{Department of Physics \& Astronomy, 
University of British Columbia, 
Vancouver, B.C., 
V6T 1Z4.  E-mail
surname@astro.ubc.ca}

\altaffiltext{2}{Rutgers University, Department of Physics and Astronomy, PO Box 849,
Piscataway, NJ 08855--0849. E-mail pryor@physics.rutgers.edu}

\altaffiltext{3}{University of Maryland, Department of Astronomy, College Park, MD
20742--2421. E-mail rabell@astro.umd.edu}

\altaffiltext{4}{University of California, Lick Observatory, Santa Cruz, CA 95064.
E-mail bolte@ucolick.org}

\altaffiltext{5}{Space Telescope Science Institute, 3700 San Martin Drive, Baltimore,
MD 21218. E-mail bond@stsci.edu}

\altaffiltext{6}{McMaster University, Department of Physics and Astronomy, Hamilton,
ON, Canada L8S 4M1. E-mail harris@physun.physics.mcmaster.ca}

\altaffiltext{7}{Dominion Astrophysical Observatory, Herzberg Institute of
Astrophysics, National Research Council, 5071 W. Saanich Road, RR5,
Victoria, B.C., Canada V8X 4M6. E-mail firstname.lastname@hia.nrc.ca}

\altaffiltext{8}{Florida Institute of Technology, Dept. Physics \& Space
Sciences, 150 W. University Blvd, Melbourne, FL 32901-6988. Email
wood@kepler.pss.fit.edu}

$\dagger$ This work was based on observations with the NASA/ESA {\it Hubble Space
Telescope}, obtained at the Space Telescope Science Institute, which
is operated by AURA, Inc., under NASA contract NAS5-26555.

\newpage

Submitted to the Astrophysical Journal, November 24, 1996.

\begin{abstract}

Using WFPC2 on the {\it Hubble Space Telescope}, we have isolated a sample of 
258 white dwarfs (WDs) in the Galactic globular cluster M4. Fields
at three radial distances from the cluster center
were observed and sizeable WD populations were found in all three. The location
of these WDs in the color-magnitude diagram, their mean mass of 0.51($ \pm 0.03$)M$_{\odot}$, and 
their luminosity function confirm
basic tenets of stellar evolution theory and
support the results from current WD cooling theory. The WDs are used to
extend the cluster main-sequence mass function upward to stars that have already
completed their nuclear evolution. The WD/red dwarf binary
frequency in M4 is investigated and found to be at most a few percent of all
the main-sequence stars.  
The most ancient WDs found are $\sim 9$ Gyr old, a level which is set
solely by the photometric limits of our data. Even though this is less than the age of M4, we discuss how these cooling WDs can eventually be used to check 
the turnoff ages of globular clusters and hence constrain the age
of the Universe.

\end{abstract}

\keywords{clusters: globular, clusters: individual M4, stars: white dwarfs}


\section{Introduction}

Stellar evolution theory predicts that all single, low-mass stars end their
lives as white dwarfs (WDs). This fundamental prediction has, however, never actually been observationally confirmed. Attempts in this direction were made by Weidemann (1989)   who compared the planetary nebulae birth rate with that of WDs, but later showed (Weidemann 1990) that not all WDs go through the planetary nebula evolutionary channel. White dwarfs in open clusters have been observed to determine ages, the upper mass limit of WD projenitors and relative birthrate statistics ({\it e.g.} Luyten \& Herbig 1960; Woolf 1974; Hartwick \& Hesser 1978; Romanishin \& Angel 1980; Anthony-Twarog 1981, 1982; Koester \& Reimers 1982, 1993, 1996; von Hippel {\it et al.} 1996), but 
the population
of WDs in any open cluster is always small.
What is required are observations of a rich stellar cluster
that penetrate faint enough so that the WD members can be found in
sufficient numbers. The number of WDs detected can then be compared with 
a theoretical prediction based on the number of bright stars in some well-defined evolutionary phase with its associated timescale, and the WD cooling time
to the limit of detection. Clearly, globular star clusters are 
an excellent testing ground for such an approach. No convincing WD sample was ever isolated in a globular cluster from ground-based
work despite several attempts (Richer 1978; Chan \& Richer 1986; 
Ortolani \& Rosino 1987; Richer \& Fahlman 1988)  
as the distances to the clusters makes the WDs faint and the 
crowding of the stellar images in their dense inner regions severe. 

With the repair of {\it HST}, crowding no longer remains as a barrier and
there are three recent reports of significant populations of WDs in
the nearest clusters (Richer {\etal} 1995, Cool {\etal} 1996, Renzini {\etal} 1996). The first of these papers announced a well-populated WD sequence in M4
and derived a preliminary luminosity function for them. Cool {\etal} 
discovered a modest population of a few dozen WDs in NGC 6397 while Renzini {\etal} used
the WD cooling sequence in NGC 6752 to derive an accurate distance to the
cluster.     
In the present
paper we discuss the largest sample of WDs yet discovered in a stellar cluster, 
that in  
the Galactic globular cluster M4. We find that
the M4 WDs possess an average mass ($0.51$M$_{\odot}$) very close to 
that expected from
stellar evolution theory  ({\S5}), and that they appear
in numbers consistent with the hypothesis that {\it all} evolving stars in
the cluster for the past 4 Gyr have terminated their lives as WDs ({\S6}). 

The oldest WDs which we observe in M4 have cooling ages of ${\sim 9}$ Gyr.
This detection limit is currently governed only by the length of the {\it HST}
exposures.  
In {\S9} we discuss whether globular cluster WDs can be detected to
cosmologically interesting ages and thus whether they can be used to check 
cluster turnoff ages and hence constrain the age of the Universe.


\section{Observations and Data Reduction}

The observations discussed in the following sections are imaging data
obtained with {\it HST} using WFPC2 in cycle 4 (GO-5461). Three fields were imaged: 
one whose PC field was within one core radius (${r_c}$) of the cluster center,
which we take as 50${\arcsec}$ 
(Djorgovski 1993); a second whose PC field was at about 1.7${r_c}$;
and a third whose PC field was near 6${r_c}$. In the inner two fields the
cluster was imaged through three filters: F336W ($U$), F555W ($V$) and F814W ($I$).
For the outer field we only exposed through the two reddest filters. The 
aim of the program was to use the long color baseline of F336W - F814W 
($U - I$) in those regions
where we expected the largest number
of WDs so that we could best define the WD cooling sequence, and to 
reach as faint as possible 
for the cooler WDs and the redder main sequence stars in the outermost uncrowded field. 
Total exposure 
times in the two inner fields were 11,800, 15,000 and 5,500 seconds in F336W,
F555W and F814W, respectively. In the 6$r_c$ field the exposures were 31,500 seconds in F555W and 5,500 seconds in F814W.   

The locations of the measured stars in these fields are illustrated
in Figure 1 where annuli boundaries at 0.5, 1.5, 2.5, 4 and 8${r_c}$ are displayed 
centered on the cluster
together with all of the stars found in each of them. In the discussion which
follows we will be referring to the stars by annulus with annulus 1 lying between 0.5 and 1.5${r_c}$, annulus 2 between 1.5 and 2.5$r_c$, annulus 3 between 2.5 and 4.0${r_c}$ and with annulus 4's PC field centered near 6.0${r_c}$. At the
time when the data were taken, it was not as yet clear whether it would be
possible to reduce the badly undersampled data from the WF cameras. Experience
has now taught us that the WF data are almost as reliable as those of the PC, 
but at the time of setting up the program, we made sure that there were
no very bright stars on the PC frames
and we let the WF cameras fall as they may without trying to control the
spacecraft's roll angle. For this reason there is
quite a bit of overlap in annulus 2, {\ie} a large fraction of these stars
were measured in two separate pointings of the telescope.

The raw {\it HST} data frames had the standard pipeline processing performed on
 them; this includes bias subtraction, correction for dark current, and
 flat-fielding. In addition to this, we flagged hot pixels and did not use them in the
 reductions, vignetted pixels were blanked out, the charge transfer efficiency
 correction was made, the decrease in F336W transmission as a function of time
 from the last decontamination was accounted for, and we corrected for nonuniform
 illumination of the chips. The photometry was carried out using ALLFRAME
 (Stetson 1987, 1994) with a quadratically-varying point-spread function (PSF).
 Complete details concerning these reductions can be found in Ibata 
{\etal} (1997).
 
This technique of PSF-fitting in 
deriving magnitudes and colors for stars requires that we are able to relate
a PSF magnitude, which is only sampled around the intensity peak, to one derived from a large aperture which envelopes the entire light from the star. Instead
of attempting to derive the aperture corrections to the PSF magnitudes directly
from the data (which are often badly corrupted by cosmic rays), we produced a
number of artificial frames (containing appropriate sky noise) which contained
realizations of the PSF free of cosmic rays and measured the corrections 
on these artificial frames. 

To permit comparison with theoretical results and with ground-based
 photometry, all of the
 data were transformed to $U$, $V$ and $I$ as defined by the Landolt standards
 (1983, 1992a,b) using the transformations discussed in Holtzman {\etal} (1995).
In addition to this transformation, we included the +0.05 mag correction which
Kelson {\etal} (1996) have shown is required for {\it HST} images which have
in excess of 160e$^-$ in each sky pixel. For sky counts less than 160e$^-$
a linear ramp
between no correction and +0.05 mags was used.
As a check on these procedures, in Figure 2 we compare photometry
of individual stars from annulus 4 obtained with the Las Campanas Observatory 2.5 m telescope (calibrated with Landolt standards) 
with that from {\it HST} transformed to ground-based $V$ and $I$, 
as described above. These diagrams show quite
satisfactory agreement between the two data sets.
There is
some evidence of a small systematic difference between the
photometric zeropoints of the two systems amounting to about 0.03 mags in $V$
and 0.01 mags in $I$. Since it is not clear whether the ground or the space
data might contain this small error, we have not forced the two sets to
agree but have adopted the calibration for the {\it HST} data as is.

A further check
on the calibration of the photometry involves a comparison between the 
$U$, $(U-I)$ CMD from the ground and from space. The ground $U$ photometry 
was obtained from CTIO (Richer \& Fahlman 1984). 
There are no stars in common
for this check (our ground-based $U$ photometry was also obtained in annulus 4, 
for which
no F336W frames were secured), so all we can compare are the resultant CMDs
(Figure 3).
The agreement between the
locations of the major sequences of the two CMDs provides confidence in
the calibration of the {\it HST} data that we have adopted.

In Table 1 we list the coordinates and photometry of the 258 WDs found in the three fields. The objects appearing
in this table all possess $\chi$-parameters (which is a measure of the goodness
of fit between the stellar profile and the point spread function) that are
$\le 2.0$ (Stetson 1987) and photometric errors in their $V$ and $I$ colors 
that are
no larger than 0.5 mags. In a few cases the $U$ errors are very much larger
than this, which reflects the faintness in this color of the cooler WDs. The column 
headings are as follows: column 1, original field number F 
(0 = core field, 1 = 1$r_c$ field, 6 = 6$r_c$ field); column 2, C the chip on
which the WD is located (1 = PC, 2, 3 and 4 the WF chips as numbered in the 
{\it HST} Handbook); 
columns 3 and 4, the X,Y pixel coordinates of the star on the chip; 
columns 5 and 6,
right ascension and declination (equinox 2000), both measured in degrees and
derived from the pointing of the spacecraft and the pixel coordinates
using the software in STSDAS;
columns 7 and 8, the $U$ mag. and error; columns 9 and 10, the
$V$ mag. and error; and columns 11 and 12, the $I$ mag. and error. The error
estimates listed are the rms frame-to-frame variations in the measured magnitudes.   


\section{The Distance and Reddening to M4}

Both the distance and reddening to M4 are somewhat controversial. This
is likely due in part to the fact that the cluster is located in the
direction of the Scorpius-Ophiuchus dark cloud, so that the reddening is
quite high. A large reddening also introduces the possibility of
differential extinction across the face of the cluster (Cudworth \& Rees 1990a).
Several recent papers (Liu \& Janes 1990, Dixon \& Longmore 1993, Vrba {\etal} 
1993, Peterson {\etal} 1995) have further suggested that the ratio of
total-to-selective absorption in the $V$-band in the direction of M4, $R_V$, 
is larger 
than normal, in the range of 3.8. We adopt this value for the discussion which
follows below. 

The cluster distance and reddening are derived simultaneously here by fitting a set of subdwarfs
to the unevolved part of the M4 main sequence. The subdwarfs are the same
sample used by Mandushev {\etal} (1996) in their study of M55 and the M4 
color-magnitude diagram (CMD) employed is the ground-based data used above to check the
calibration of the {\it HST} photometry. We used these data instead of the
{\it HST} results as the ground-based data contain brighter stars which better match the subdwarf sample. The stars in the upper part of the unevolved main sequence are largely saturated on the {\it HST} frames.
   
The theoretical models of Bertelli {\it et al.} (1994) were used differentially
to adjust
the subdwarf colors to account
for their metallicity difference with respect to M4's, which is taken to be $[Fe/H] = -1.3$
(Djorgovski 1993). The best fit between 
this empirical
population II main sequence and the fiducial sequence of the cluster is
shown in Figure 4. Contours of constant $\chi^2$, the goodness of fit
between the M4 fiducial sequence and the subdwarfs for different reddening 
and distance moduli are illustrated in 
Figure 5. It is clear from the diagram that this main-sequence
fitting technique is capable
of constraining the reddening reasonably well, but does a less
accurate job of determining the best estimate of the apparent distance modulus.
 
The parameter values that minimize
$\chi^2$ are $(m-M)_V = 12.51(\pm 0.09)$ and $E(V-I) = 0.51(\pm 0.02)$. 
We will use these values for the rest of this paper. 
With $R_V = 3.8$,
we can then derive that $E(B-V) = 0.35$, $A_V = 1.32$ and, thus, 
that the distance
to M4 is 1.73 kpc. The expressions developed in Cardelli {\etal} (1989) were
used to determine $E(B-V)$ from $E(V-I)$ with our non-standard value of $R_V$.
The present value for $E(B-V)$ is in excellent agreement with direct
measurements of this quantity in M4 ({\eg}, Richer \& Fahlman 1984; Dixon \& Longmore 1993). Moreover, the
rather small cluster distance is in almost perfect accord with the
astrometrically determined distance of $1.72 \pm 0.14$ kpc (Peterson {\etal} 1995) and the 1.73 kpc derived from the Baade-Wesselink analysis of M4 RR Lyrae
stars (Liu \& Janes 1990).

\section{The White Dwarf Cooling Sequence}

Since all single stars currently completing their nuclear evolution in a
globular cluster end up as WDs, and since this statement has been true
for many billions of years now, we expect a substantial population of WDs
to be present in M4. Thus our first aim is to demonstrate whether WDs
are indeed present and whether they are located in the expected regions
of the CMD. Section 6 of this paper will
address the question of whether the WDs are present in the expected numbers. 

The left-hand panel of Figure 6 displays the
$M_U$, $(U-I)_O$
CMD for all stars possessing $U$ photometry (annuli 1-3) with errors in
each color that are less than 0.25 mag and $\chi < 1.5$
(Stetson 1987). The appearance of this CMD differs somewhat 
from our earlier version (Richer {\etal} 1995) with these same data, as we have
reprocessed the images with a smaller fitting radius which is less capable of measuring saturated stars, but does a better job with the faintest objects.   
The long color baseline was chosen to make the WD cooling sequence
as distinct as possible from that of the main sequence.  
Two well-defined
sequences are present in this diagram, the main sequence of the cluster and
a roughly parallel sequence containing 109 stars about five magnitudes bluer in $(U-I)$ 
that begins
at $U$ magnitude near 22 ($M_U = 9$) and continues to the limit of the data at $M_U = 12.5$ ($U = 25.5$). 

The scatter of stars below the main sequence is contributed by the
Galactic bulge/spheroid as M4 is roughly in the Galactic center direction.
Additional discussion of this component of the CMD can be found in Fahlman {\etal} (1996a, 1996b, 1997).
The locus of the bluest stars
is certainly what a WD cooling sequence is expected to look like. We confirm
that suspicion by replotting the data and
overlaying a theoretical cooling curve for 0.5M$_{\odot}$ hydrogen-rich 
(DA) WDs (Figure 6, right-hand panel). The theoretical
locus was derived from the evolutionary models of Wood (1995) and the DA 
atmospheres of Bergeron {\etal} (1995). Wood's models provide the WD mass and radius at a given $T_{eff}$ or age. From these numbers, the gravity of the star was calculated. We then interpolated in the table of atmospheric colors of Bergeron {\etal} to obtain colors at the appropriate gravity and $T_{eff}$. The agreement between the theoretical and observed loci
makes it seem incontrovertible that the
sequence of blue objects is indeed the cluster WD cooling sequence. 

We note
in passing the hot (27,000 K) WD at
$M_U = 8.5$; according to the evolutionary models of Wood (1995), a WD 
at this temperature is only 13 million years old. Neutrino losses, while
somewhat less than the photon luminosity of the star, are still an important energy loss mechanism for this WD. This star is in the region of the WD cooling
sequence where variable DB stars (He-rich atmospheres) are found, and is bright enough ($V = 22.08$)
that it could
be monitored from the ground with a 4-m class telescope. In Figure 7 we
present a finding chart for this object. The ZZ Ceti instability
strip occurs near $(U-I) = -0.4$ or $(V-I) = 0.0$ which corresponds to 
$V \sim 24.1$ in M4. This is probably too faint for a ground-based search for
variability with existing telescopes. 

As a further test of the location of WDs in the cluster CMD, and inferentially
as a check on the value of $R_V$ and the reddening used, we display in
Figure 8 the $M_V$, $(V-I)$ CMD for data from the  four annuli. Again, 
these data
have been selected to have errors in their magnitudes of $\le 0.25$ and 
$\chi$ values $\le 1.5$. In the companion diagram we overlay 
the theoretical cooling curve for
0.5M$_{\odot}$ DA WDs in these colors.
As was evident in Figure 6, the agreement between the
observed and theoretical loci is superb. 

\section{The White Dwarf Masses}

The mean mass and the distribution of masses among the cluster WDs provide
sensitive tests of stellar evolution theory and constrain the
fraction of binaries in the cluster. For example, the mean WD mass is set by
the electron degenerate core mass at the termination of the asymptotic
giant-branch 
phase of evolution and is predicted to be 0.53($\pm0.02$)M$_{\odot}$ for stars
currently terminating their evolution in a typical globular cluster (Renzini {\etal} 1996, Renzini \& Fusi Pecci 1988).
WDs with masses higher 
than this mean value could have originated from more massive progenitors,
such as blue stragglers (BS), some of which are thought to be merged binaries.
Good reading on BSs in globular clusters can be found in articles by
Sarajedini, and Bailyn and Mader in "Blue Stragglers" (1993). 
Lower mass
WDs could have evolved from mass transfer binaries in which a red giant with a predominantly helium core transfers mass to a companion and interrupts its
evolution prematurely. Thus, the relative number of WDs with masses 
significantly different than the predicted mean of 0.53M$_{\odot}$ 
may be related to the fraction
of binaries present in the cluster. An estimate of the binary fraction 
involving non-interacting WD/red dwarf pairs will be given in 
$\S 7$. 

We illustrate how low- or high-mass WDs could be
recognized in the CMD by replotting in the left-hand panel of Figure 9 the $M_U$, $(U-I)$ CMD of
Figure 6 together
with cooling curves for 0.4 to 1.0M$_{\odot}$ WDs in increments 
of $0.2$M$_{\odot}$. These cooling sequences are for
pure carbon core evolutionary models with helium layers having 1$\%$ of the mass of
the star and thick hydrogen layers of 0.01$\%$ by mass. The atmospheric colors are for DA models and are taken from Bergeron {\etal} (1995). 
The core composition makes little
difference to the location of these cooling curves in the CMD 
(only the cooling times to
a given luminosity are affected), but the composition of the atmosphere does. We discuss this latter
point below. 

To estimate
the mean of the WD mass distribution, we fit a grid of cooling curves for DA WDs to our sample of WDs using the standard
$\chi^2$ goodness-of-fit statistic.  The observed white dwarf colors
were weighted by $w = 1 / \sigma_s^2$, where $\sigma_s$ is the
frame-to-frame scatter in the magnitudes determined by ALLFRAME. The M$_{WD} = 0.51$M$_{\odot}$ DA cooling curve
gave the best fit to the data in both the $U,U-I$ and $V,V-I$ planes.
In order to determine the uncertainty in this mass estimate we did a
series of 1001 bootstrap resamplings of the data using standard
bootstrapping techniques ({\it e.g.} Hill 1986, Feigelson \& Babu 1992).
That is, we took the original sample of 109 WDs (the number of WDs in Figure 6) and randomly
picked, with replacement, 109 WDs from this sample.  We then
did a $\chi^2$ fit to the resampled data in exactly the same manner
as we did to the original data to determine the best-fitting DA
cooling curve mass.  The distribution of masses obtained from 1001
bootstrap resamplings was consistent with a Gaussian distribution.  The
mass distributions obtained from bootstrapping the data in each plane
had means of $\overline M_{WD} = 0.507 \pm 0.025 
M_{\odot}$ in the $U,U-I$ plane and $\overline M_{WD} = 0.510
\pm 0.023 M_{\odot}$ in the $V,V-I$ plane.  The quoted
uncertainties are one standard deviation.  These results suggest that
our data are consistent with the WDs in M4 having a single
mass of 0.51M$_{\odot}$ with a 1-$\sigma$ uncertainty of 0.03M$_{\odot}$.
In contrast to the mean mass for these globular cluster WDs, the mean of the field WD mass
distribution is somewhat higher at 
0.59M$_{\odot}$ 
(Bragglia, Renzini \& Bergeron 1995). This is not surprising considering
that the field sample is likely to have evolved, on average, from more 
massive progenitors which leave higher mass remnants.

The above result for the mean WD mass 
is likely to be somewhat in error as the 
possible presence of non-DA stars was not considered. The effect of a
component of DB WDs, for example, on the observed WD sequence can be judged
from the right-hand panel of Figure 9, where we illustrate the locus of 0.5M$_{\odot}$ cooling DA 
and DB stars.
If DB WDs are in fact present in the cluster with the same ratio as in the field
where 
about 20\% of field WDs are of type DB (Sion 1986), then they will produce an apparent 
tail to higher masses if the WDs are analysed under the assumption of a pure DA sample. This is not
a problem above a temperature
of about 30,000 K where there are no DB WDs but, unfortunately, no
WDs at this high a temperature are present in our sample. However, as can be seen in Figure 9, DAs and
DBs are almost indistinguishable in the $V, V-I$
plane.  The separation between the DA and DB
curves is of order or smaller than the dispersion along the cooling sequence. 
The good agreement between the mean masses derived in the two CMDs then 
argues that DBs have not been a serious problem.

Had we chosen, instead,  to adopt a mass for the M4 WDs based on some initial-final mass
relation, we could then have used the WDs to
derive an accurate cluster distance as Renzini {\etal} (1996) 
have done for NGC 6752.
If we choose, as they did, a mass of 0.53$M_{\odot}$ for the WDs in a globular
cluster and fit a cooling sequence of this mass to the observed WDs, we would then derive a distance modulus to M4 of $(m-M)_V = 12.60$. 
This is to be compared with $12.51$ derived from the main sequence.

\section{The White Dwarf Luminosity Function}

Thus far we have shown that WDs are present in globular
clusters and that they are located in the expected regions of the CMD, or,
alternatively, that they possess about the masses predicted by theory.  
With the luminosity
function we explore whether WDs are present in the expected
numbers. This provides a sensitive test of the rate at which the WDs are
cooling and thus of the evolutionary models, which, except for the
preliminary analysis in Richer {\etal} (1995), have never actually
been tested against a sequence of cooling WDs. The only other possible, similarly
sensitive, test is to look for photometric period changes
in variable WDs caused by cooling (Kepler {\etal} 1991, Bradley {\etal} 1992). 
These effects are sufficiently small that definitive results are still 
lacking for variable  DA ($T_{\rm eff}\sim12,000$~K) and DB ($T_{\rm
eff}\sim27,000$~K) WDs.  
To test the theory in detail,   
we develop the WD cumulative luminosity function
(CLF) and compare it with theoretical luminosity functions constructed with 
three different core compositions, namely, pure carbon, mixed carbon and
oxygen, and pure oxygen.

The CLF is built by counting WDs as a function of magnitude in each
annulus and correcting the star counts for incompleteness. It is clear from
Figures 6 and 8 that no large corrections are required for either foreground or
background objects. There may be a few contaminating WDs contributed from the
Galactic bulge region. These objects should begin at an absolute $V$-magnitude
near $+13$ in Figure 8 (the Galactic bulge possesses a distance modulus that is
about 3 magnitudes larger than that of M4), but we estimate that their numbers must be no more
than a few percent of the cluster WDs given the paucity of 
red giants contributed by
the bulge in this diagram.
Background blue galaxies that appear stellar are also rare.
In the Hubble Deep Field, Reid {\etal} (1997) find two blue $(V-I \le 1.0$) stellar objects, both
with $V < 21$ and thus brighter than the WD sequence seen in 
Figure 8. In
addition they find three $V \sim 27$ blue objects which {\it might} 
be stellar. Because of the apparently small contribution to the cluster WD sequence,
we have chosen to ignore corrections for either foreground 
or background objects.

The incompleteness corrections to the counts were obtained
by adding stars with WD colors and magnitudes along a dispersionless
distribution (see Figure 10) to random locations in the frames
and rereducing the frames to obtain the recovery statistics for the WDs. The WDs
were added in small numbers (typically 150 per frame) so that they would
not seriously
affect the crowding statistics in the images. In total, almost 16,000
artificial WDs
were added in more than 100 trials and the input and output sequences are illustrated in Figure 10. 
The CLF was constructed without binning as each counted star was simply 
replaced by ${\it n}$ stars, 
where ${\it n}$ is the inverse of the completeness fraction at that magnitude
along the WD cooling sequence. Figures 11a (for the WF chips) and 11b 
(PC chips) display these incompleteness
corrections in the four annuli in $V$ for stars that are found on {\it both} $V$ and $I$. In the analysis which follows, we set the magnitude limit for
inclusion in the WD luminosity function to the point where the completeness 
in the sample drops to 30\%. Determining these corrections in
ALLFRAME is extremely cpu intensive as the WDs are added to ${\it all}$ of the individual frames which are then processed exactly as the 
original frames. About 6 cpu months on a SPARC 20 were 
required to complete this analysis. Further details relating to the procedures
employed can be found in Ibata {\etal} (1997).

Figure 12 illustrates the observed CLF for the WDs in the four annuli. The data
extend to deeper magnitudes in more distant annuli mainly due to the
decreasing effects
of crowding and scattered light from saturated stars. Interpreting Figure 12 requires comparing the observed CLFs with those derived by combining theory and the number of WD progenitors in each field. However, what can be said is that the slope of the differential luminosity function (d[log $N_{WD}$]/d[$M_V$]) for the cluster WDs in annulus 4, $0.29\pm0.07$, is consistent with the $0.39\pm0.05$ found for a field sample (Winget {\etal} 1987, Liebert {\etal} 1988, Oswalt {\etal} 1996). 

We compare the observed CLF with theory by initially making the assumption that
the number of stars in any region of the cluster is
conserved as they move through their post-main-sequence
phases of evolution. This does not allow for evaporation or tidal stripping 
of stars 
from the cluster or, more importantly, does not take into consideration mass segregation (see Fahlman {\etal} 1996b for a detailed discussion of this point). The hypothesis of a conserved number of stars will become increasingly
less valid the
older the WD because the cooling time of the star can become longer than the relaxation time of the cluster. However, the obvious first step is to
compare the number of WDs present in some field with that of a tracer population with a well-defined mass and lifetime. With such an approach,
the effect of mass segregation becomes strictly a differential one due to the difference in mass between the WDs and the tracer. A convenient stellar population for this 
purpose is the cluster horizontal branch (HB) stars.    

If there is no differential mass segregation between HB and WD stars, 
then the numbers of stars in the two 
post-main sequence phases will be strictly
determined by the ratio of the lifetimes of those phases. This yields 
\begin{equation}
N_{WD}(<M_V)=N_{HB} \cdot \left({t_{cool}(M_V)\over t_{HB}}\right)
\end{equation}
where $N_{WD}(<M_V)$ is the number of WDs brighter than absolute magnitude
$M_V$, $N_{HB}$ is the number of HB stars, $t_{cool}$ the WD cooling time to reach $M_V$, and $t_{HB}$ the lifetime
of a star on the HB, which we take as $10^8$ yr (Renzini 1977, Dorman 1992). All
of the WD theory is in $t_{cool}$, which comes from Wood's (1995) models 
together with the model atmospheres
of Bergeron {\etal} (1995). $N_{HB}$, the number of HB stars within the area of each annulus covered by our {\it HST} observations, was estimated 
from ALLFRAME photometry of a 300 sec $B$-image
and a 160 sec $V$-image taken (by C. Pryor) with a Tektronix 2k CCD on the 0.9-m
telescope at Kitt Peak. Since the number of HB stars is always small (there are only $\sim 150$ HB stars in the entire cluster), we measured the numbers
of giant branch (GB) stars with $V < 16$ and HB stars at different distances from the cluster center and used the average ratio of 0.135 HB stars/GB star to set the HB number. Both the GB and HB stars are bright enough that their 
numbers have
not been affected significantly by incompleteness. This ratio represents a global average as it is determined from data extending well past the
half mass radius for which the relaxation time is too long to have caused
significant mass segregation between the HB and GB stars. 
Because we do use the number of GB stars to represent the HB population, our HB stars are distributed in the cluster {\it as if} they possessed masses of $0.81M_{\odot}$,
which is the mass of turnoff stars in M4 (Bergbusch \& VandenBerg 1992).

We cannot expect that
agreement between the theory and observations will be perfect as the sample 
of normalizing stars is not large, differential mass
segregation between the ($0.81M_{\odot}$) HB stars and the ($0.51M_{\odot}$) WDs has not been accounted for, and the WD population is modest. In the first comparison between theory and observations (Figure 13)
we use models with pure carbon core
interiors together with DA atmospheres. While the similarity between theory and observation is apparent, there remains a systematic trend in the normalization between the two. The normalization is too high in the inner annuli (too many HB stars compared to the number of WDs) and too low in the outer ones (too few HB stars). This is exactly what is expected from mass segregation (see Fahlman {\etal} 1996b). 

We correct for differential mass segregation by fitting  the projected density profile of the GB stars 
to the GB mass profile in a multi-mass Michie-King (1966) model. This fit
yields the total number of such stars in the cluster. We then calculate
the projected density profile of a $0.51M_{\odot}$ component in the
model that has the same total number of stars.  This assumes that the
WDs have come into equilibrium (as defined by a King model)
at all radii.  This is somewhat extreme, as the relaxation time at
large radii can exceed the age of the cluster.
Using the surface density of the $0.51M_{\odot}$ component, we then predict
the number of HB star progenitors for each of the four WD CLFs by
multiplying by the area observed and HB/GB = 0.135. This result is only very weakly dependent on the slope of the cluster mass function used in the model; for a slope of $x = 0.7$ (Salpeter = 1.35), we obtain the normalizations indicated in Figure 14. Here the agreement is much improved over Figure 13: only
the CLF in annulus 2 appears to deviate significantly from the theoretical
expectations. However, even in this case the difference at the faint end
only amounts to about a 1.5-$\sigma$ deviation, a difference that is seen by chance 13\% of the time. Figure 14 illustrates that when the 
observed amount of mass segregation between the WDs and the GB stars is taken
into account, the CLFs  
provide a strong positive test of the correctness of current WD cooling theory.   

The analysis presented above is strictly correct only for WD cooling times,  
$t_{cool}$, that
are small compared to the age of the cluster. If the cooling times are an appreciable fraction of the cluster age, the slope of the cluster initial
mass function (IMF) and the relation between stellar mass and main sequence lifetime
become important. For example, if we choose a global power-law IMF
of the
form $n(m) \propto m^{-(1 + x)}$ and the current main-sequence lifetime as a function
of mass is represented by $t_{MS} \propto m^{-\gamma}$, then in the limit of long cooling times ($t_{cool} \rightarrow$ age of the cluster),  
\begin{equation}
N_{WD}(<M_V) = N_{HB} \cdot \left({t_{cool}(M_V)\over t_{HB}}\right) \cdot \left({{\gamma}\over x}\right). 
\end{equation}
Reasonable values for $x$ and $\gamma$ are 0.7 and 2.5, respectively, so that
the expected number of faint WDs can potentially be increased by up to a 
factor of three. 

We can repeat the analysis displayed in Figure 14 with core compositions
consisting of a mixture of carbon and oxygen or of pure oxygen (evolutionary
models with these compositions have been calculated by Wood 1995), and ask if
the comparison with theory is similarly satisfactory. WDs will cool at
different rates with these varying core compositions since the heat capacity
of the atoms differ. For example, a WD with a pure oxygen core is
younger than one with a pure carbon core at a given luminosity. This, then, is potentially an effective way of determining the
core composition of these objects. In Figure 15 we compare our observations
to Wood's models for pure oxygen, which is a more dramatically
different core composition than the carbon-oxygen mixture. As can 
be seen, 
the results 
differ little from those in Figure 14: 
small changes in the normalization will make the observations fit the data
as well as they do for the carbon-core models. The conclusion must then be that the sample of WDs itself and/or the number
of normalizing HB stars remain too small, or that 
the test is simply too insensitive to use as a core chemical composition indicator.  

Theory could potentially shed some light on the WD composition. Post-asymptotic giant branch evolutionary models of several
researchers have agreed for a number of years on the general trends in the
$\rm C/O$ profiles of the WD remnants they produce: an $\rm O$-rich inner core, surrounded by a
large transition zone which tails off to a nearly-pure $\rm C$ outer core ({\it cf.}
Mazzitelli \& D'Antona 1986 with 
Bl\"ocker 1995).  However, these profiles
are sensitive to the $\rm{}^{12}C(\alpha,\gamma)^{16}O$ reaction rate,
which has proved difficult both to calculate and to measure (Azuma {\it et al.} 1994). This
inability to distinguish cleanly the chemical make-up of the core has a modest 
effect
on using the WD CLF as an astronomical clock.  The
resulting uncertainties in composition translate to $\sim$5\%
uncertainty in the ages.
       
\section{Binaries Among the White Dwarfs}

Are a significant fraction of WDs
tied up in binary systems with main sequence stars? The binary
fraction is of general interest in dynamical studies of globular clusters
since heating from
close binaries is expected to inhibit, delay, or even reverse core collapse in 
these systems
(Hut {\etal} 1992). Detailed discussions of the fraction of binaries among 
M4 main-sequence stars can be found in Fahlman {\etal} (1997), Pryor {\etal}
(1996) and C\^{o}t\'{e} \& Fischer (1996). 

We can use the color-color diagram of the cluster to
investigate the frequency of binaries involving stars of very different colors
(see Richer {\etal} 1996b). The system of a red dwarf star in a binary with 
a WD will
have a rather peculiar color. In Figure 16 we plot
the color-color diagram
for all of the stars measured in three colors (annuli 1-3), and we 
include the results from
a simulation that draws 100 stars at random from the WD sequence and the main
sequence and then determines the color of the resulting object. Main sequence
stars were chosen only over the same magnitude range as the WDs so no very
bright red dwarfs went into the simulation.  
The one realization of this shown in Figure 16 (plotted as triangles) 
illustrates the 
location of
WD/red dwarf binaries in the color-color plane. About 50\% of the simulated
binaries lie very close to the main sequence as in these cases the color 
is dominated by that of the red dwarf. In the regime where the simulated binaries are
well removed from the main sequence, examination of Figure 16 indicates 
that $\le 15$ 
of the M4 stars are good candidates for such binaries. Given the
crowding statistics on the frames, as judged by the artificial star tests, very
few if any of these should be optical ({\ie}, non-physical) systems. Thus, with more than 1000 main
sequence stars in Figure 16 covering the same magnitude range as the WDs, less than 2\% of all such objects appear to
be involved in binary systems with a WD. This number could be higher by about
a factor of two or so in order to account for systems lying too close to
the main sequence to be distinguished from single stars.

 \section{Extending the Cluster Main Sequence Mass Function}

Up to what mass does the IMF of a globular cluster extend? 
The answer is currently unknown, but it is certain to extend at least up 
to main sequence masses 
required to produce neutron stars as, at last count, 34 pulsars had been 
found in
globular clusters (see e.g. Lyne {\etal} 1996).  It is important
to establish this upper limit as the dynamical history of the cluster is to a great extent controlled by its IMF. Possible self-contamination of the proto-cluster cloud
with metals will also depend on the number of massive stars originally formed,
with differing star formation scenarios in the early Universe predicting diverse
values for the massive star fraction.  

According to the M4 photometry of Richer \& Fahlman (1984) and the models of Bergbusch \&
VandenBerg (1992), turnoff stars in M4 possess masses of $\sim 0.81 M_{\odot}$.  Thus from an examination of the mass function of present-day 
hydrogen-burning stars in this cluster, we learn little about the massive stars
initially present. The cluster WDs, having evolved
from stars more massive than the current 
cluster turnoff, provide the potential of extending the mass function
to heavier objects: the older the WD, the more massive 
its progenitor.
 
As we will see in {\S 9}, the oldest WDs currently observed in the cluster have 
ages $\sim 9$ Gyr. 
If the cluster is 15 Gyr old, then these WDs evolved from stars with main-sequence lifetimes of $\sim 6$ Gyr. Such stars have initial masses of $\sim 1.2 M_{\odot}$, 
so the current data potentially allow us to extend the cluster mass function by
about 0.4$M_{\odot}$. However, the reality is that we are not confident
in the completeness corrections when they fall below 30\%, which 
limits the oldest WDs for which we have convincing statistics 
to an age of only 3.7 Gyr. The progenitor
masses in this case are $\sim 0.9M_{\odot}$. Hence we can confidently extend
the M4 mass function by only 0.1$M_{\odot}$. We carried out this exercise and found that the extension of the M4 mass function derived by Fahlman {\etal} (1997) up to 0.9$M_{\odot}$ appears to be 
rather flat ($x = -1$). However, this likely
does {\it not} represent the IMF, as 
neither the loss of stars from the cluster nor mass segregation have been considered here. These are likely to have an important influence on the cluster
mass function as its small $w$ velocity (Cudworth \& Rees 1990b) suggests
that M4 spends most of its orbit near the Galactic disk where shocking and tidal effects
can efficiently remove low mass stars from it. 

There is, however, the potential of using the WD CLF to explore the IMF of a globular cluster to much higher masses. The dependence of $N_{WD}(< M_V)$ on
the IMF slope $x$ is explicitly given in equation 3, which is correct to 
second order in the WD cooling times and for which
equations 1 and 2 are limiting cases: 
\begin{equation}
N_{WD}(<M_V) = N_{HB} \cdot \left({t_{cool}(M_V)\over t_{HB}}\right) \cdot \left (1 + {t_{cool}({\gamma} - x)\over 2{\gamma}t_O}\right). 
\end{equation}
Here $t_O$ is the current age of the cluster. 
While this expression is
dependent on the assumption of star conservation, stellar losses and mass
segregation could, in principle, be accounted for. Deviations from a power-law slope in the CLF would then contain information on the cluster IMF. We explicitly demonstrate the effect on the WD CLF of including the mass spectrum and variation in main sequence lifetime with mass by replotting in Figure 17 the
data from annulus 4 and using equation 3 to represent the theory. The effect of
including the cluster mass function slope is to decrease the number of WDs (as  
$x$ is generally positive this produces fewer high mass stars that eventually become WDs). The inclusion of the main sequence lifetimes acts in the opposite sense, as the more massive stars spend a shorter time on the
main sequence and hence produce WDs more quickly. For $x = +2.5$ and ${\gamma} =
 2.5$ the two contributions cancel each other. Figure 17
illustrates that over the extreme range of $x$ for fixed ${\gamma}$, the expected number of WDs to an M$_V = +16$ will vary by about 40\%. Taking Figure 17 at face value, it appears that under the unlikely assumption of no loss of stars from the cluster, the IMF of M4 must be rather steep if ${\gamma}$ is in the reasonable range of 2.5. Allowing for the depletion of old WDs
through some star-loss mechanisms suggests that a flatter IMF is allowed.

\section{White Dwarfs and the Age of the Universe} 

There is currently an unexplained difference in the ages of the Universe derived from the expansion and from its oldest datable component, the globular clusters. With $H_O = 70$ km/sec/Mpc (a median value from the recent Space Telescope
Science Institute  
workshop on the distance scale), standard inflationary cosmology 
($\Omega = 1, \Lambda = 0$) yields 9.5 Gyr for the expansion age, and hence $\lesssim9$ Gyr for the age of the oldest globular clusters.
By contrast, globular cluster ages, derived via the 
luminosity of the main-sequence turnoff, are in the range of 14-16 Gyr (Richer {\etal} 1996a, Chaboyer {\etal} 1996, Bolte \& Hogan 1995). While it is possible to reconcile these two ages with non-standard cosmologies ({\eg},  
$\Lambda \neq 0$), such a solution will not be compelling until all reasonable avenues to check these ages are explored. A new approach to the globular cluster ages can come from the ages of their coolest WDs. The age of
a WD is directly related to its luminosity, and the time for a typical WD to cool to invisibility exceeds any reasonable estimate for the age of the Universe. Thus, a cutoff in the WD luminosity function at low luminosity will provide the time since the formation of the first cluster WDs, and hence a lower
limit to the age of the Universe.

The accuracy of WD cooling theory will be an important issue in this discussion.
To the confidence
expressed in $\S 7$ in the current models, we can add the following points. 
The most ancient WDs included in  
annulus 4 of Figure 14 are almost 4 Gyr old. Older WDs are measured but are not included as the incompleteness corrections for such faint objects are too insecure. Crystallization
of the core of a 0.5M$_{\odot}$WD begins
when it is only 3 Gyr old, so the data demonstrate that the models seem to be
reliable even into this regime where the physics of the interior is
more complex. Further, in the derivation of the age of the disk of the Galaxy using WDs (Winget {\it et al.} 1987),  much cooler WDs were used. In 
these older stars almost the entire core
of the WD has crystallized, yet the slope of the WD luminosity function down to
at least 9 Gyr agreed well with the theory.

To explore the ages of the WDs currently seen in M4, we plot in Figure 18
the $V$ and $I$ data selected 
to have errors in their magnitudes of $\le 0.5$ and 
$\chi$ values $\le 2.0$. These are more liberal criteria so this CMD
will contain fainter and more poorly measured stars than the one in Figure 8.   
The
right-hand panel of the Figure overlays a 
theoretical cooling curve for 0.5$M_{\odot}$ carbon core DA WDs 
that extends to an age of 13 Gyr.
The most ancient WDs that are currently seen in M4 are $\sim 9$ Gyr old.
If we take this as a strict lower limit to the age of the Universe then
{\it this result, by itself, requires $H_O < 73$ in standard cosmological 
models with $\Omega = 1$.} These oldest WDs are within about 1 Gyr of 
the most ancient ones
currently known in the disk of the Galaxy (Oswalt {\etal} 1996, Ruiz {\etal} 1995, Winget {\etal} 1987).  

The future use of WDs in probing cosmological models will come from data 
sets which
penetrate to equivalent $V$ magnitudes in M4 of 30 or more ($M_V \ge 17.5)$.
At this level, the WDs are
12 Gyr or older and begin to provide interesting constraints. Photometry at
such faint magnitudes can be obtained in about 100 orbits with the
current instrumentation on {\it HST}, and in significantly less time when the 
Advanced Camera is available after 1999.  It may also possibly 
be obtained in the infrared with
ground-based 8-10 meter class telescopes equipped with adaptive optics.



\acknowledgments

Some observations discussed in this paper
were secured at Kitt Peak National Observatory, which is operated
by AURA, Inc., under contract to the National Science Foundation.  
The research of HBR, GGF and WEH is supported in part through grants from
the Natural Sciences and Engineering Research Council of Canada, while that
of RAB, MB, HEB and CP is provided by NASA through grant number
GO-05461.01-93A from the Space Telescope Science Institute, which is operated by the Associated Universities for Reasearch in Astronomy, Inc., under NASA
contract NAS5-26555. The research of MAW is supported in part by the NSF 
(grant AST-9217988) and 
the NASA Astrophysics Theory Program (grant NAG 5-3103). RAI expresses
thanks to the Killam Foundation of Canada for funding.    
\newpage
	

\clearpage


\null
\bigskip
\centerline{\bf FIGURE CAPTIONS}
\medskip

{\sc Fig}.\ 1.--- The location of all the stars measured in M4 with respect
to the cluster center. The 5 radial distances indicated are at 0.5, 1.5, 2.5, 
4.0 and 8$r_c$ where 1$r_c$ is taken to be 50{\arcsec}. The decrease in the surface density of stars with increasing distance from the cluster center can be clearly seen in this diagram. 
  
{\sc Fig}.\ 2.--- A comparison between ground-based $V$ and $I$ photometry from
the Las Campanas 2.5 meter telescope with that from {\it HST}. All the stars
shown are in annulus 4.
 
{\sc Fig}.\ 3.--- The $U$, $(U-V)$ M4 fiducial main sequence based on Las 
Campanas and CTIO photometry (dashed line) compared with that from {\it HST} (solid line).

{\sc Fig}.\ 4.--- Best fit of a sample of subdwarfs (Mandushev {\etal} 1996)
to the ground-based fiducial main-sequence of M4 (solid line). The subdwarf colors were
adjusted to simulate stars with the same metal abundance 
as that of the cluster, [Fe/H] = -1.33. 

{\sc Fig}.\ 5.--- Contours of equal $\chi^2$ representing the goodness-of-fit
between the subdwarfs and the M4 main-sequence fiducial. The contour interval  is $1.1$ in $log (\chi^2/\eta)$, where $\eta$ is the number of degrees of freedom (8 in our case). Fitting the subdwarfs
in this manner clearly constrains the cluster reddening rather well, but does
a less good job of constraining the cluster distance modulus. The minimum
$\chi^2$, indicated by the plus sign, is obtained for $E(V-I) = 0.51$ and $(m-M)_V = 12.51$, which are the values we adopt for M4. 

{\sc Fig}.\ 6.--- The $M_U$, $(U-I)$ CMD for stars in the inner 3 annuli. All
stars present in this diagram have $\chi < 1.5$ (a measure of the
goodness-of-fit between the star and the point spread function [Stetson 1987])
and errors in photometry (judged from measurement of the same
star on numerous individual frames) less than 0.25 mags. The right hand panel
plots the same data but includes a $0.5 M_{\odot}$ hydrogen-rich WD cooling
sequence derived as described in the text. 

{\sc Fig}.\ 7.--- Finding chart for the hot WD found in M4. The axes in the
plot are in arc seconds and north and east are
indicated. Coordinates (2000.0) for the WD are $RA = 16^h 23^m 38\fs66, DEC = -26\arcdeg 32\arcmin 10\farcs90$. Pixel coordinates on our frame 
are (x,y) = (310.042, 375.391). 
 
{\sc Fig}.\ 8.--- The $M_V$, $(V-I)$ CMD for stars in all four annuli. All
stars present in this diagram were selected as in Fig. 6. 
The right-hand panel
plots the same data but includes a $0.5 M_{\odot}$ hydrogen-rich WD cooling
sequence derived as described in the text.

{\sc Fig}.\ 9.--- As in Figure 6, but in the left hand panel a sequence of 
DA WD cooling
curves have been overlaid. This sequence runs from $0.4M_{\odot}$ 
to $1.0M_{\odot}$ in increments of $0.2M_{\odot}$ and can be used to
determine the mean mass of the cluster WDs. In the companion panel, $0.5M_{\odot}$ DA and DB cooling curves illustrate
how an admixture of WD types can potentially affect the mass estimates.

{\sc Fig}.\ 10.--- The left panel is the dispersionless sequence
of stars that was input into the M4 data frames in groups of 150. These stars
were then photometered exactly as the real stars. Note that both
WDs and main sequence stars were added; however, here we
are concerned only with the WDs. The right panel shows
the output of this exercise. The recovery statistics of these added stars are
then used to estimate both the completeness of the counted stars as well as
their uncertainties.

{\sc Fig}.\ 11(a).--- The completeness percentages in $V$ of added WDs 
on the WF chips in the four 
annuli. These numbers are for stars with WD colors, thus the percent complete represents the probability of
recovering an input star at the indicated $V$ magnitude with WD colors on
{\it both} the $V$ and $I$ frames. 

{\sc Fig}.\ 11(b).--- Same as (a), except for the PC chips.  

{\sc Fig}.\ 12.--- The cumulative WD luminosity function (CLF) in the four annuli.
The CLF in the annuli more distant from the cluster center penetrate to
fainter magnitudes
mainly due to the effects of their being fewer saturated stars present
in the frames. 

{\sc Fig}.\ 13.--- As in Figure 12, except that a theoretical CLF (dotted line) is overlaid
on the data. In this case, the theoretical sequence is for DA WDs with pure
carbon cores. The normalization of the theoretical CLF is expressed in terms of 
the number of HB stars, $N_{HB}$, expected in the area of the annulus 
covered by the
{\it HST} observations. Theoretical CLFs derived using the ${\pm}2\sigma$ errors on $N_{HB}$ are plotted as dashed lines,  
clearly illustrating that the differences seen between the observations and
the theory are highly significant. Mass segregation {\it has not} been accounted for in this diagram.

{\sc Fig}.\ 14.---As in Figure 13, except that mass segregation {\it has} been accounted for in the normalization of the theoretical CLF with the HB stars. This procedure is described in detail in the text.  

{\sc Fig}.\ 15.--- As in Figure 14 except that the CLF is for
DA WDs with pure oxygen cores. Note that the fit between the observed and
theoretical CLFs differ little from those in Figure 14 and that a small adjustment in
the number of HB stars will make the comparison as good as that for the pure
carbon core models. This demonstrates that the core composition of globular cluster WDs is unlikely to be determined with this technique.  

{\sc Fig}.\ 16.--- The $(U-V)$, $(V-I)$ color-color diagram for stars in the inner three
annuli wherein F336W exposures were obtained. No reddening corrections were
applied. The cluster WDs, plotted as 
plus signs,  are the objects
with $(U-V) < 0.7$. Main-sequence stars are plotted with small
dots and the location of 100 artificial WD/red dwarf binaries are shown
as filled triangles. A few M4 objects
are possible candidates for such binary systems, but when the probability of optical binaries is included, the number is
small.

{\sc Fig}.\ 17.--- The observed WD CLF in annulus 4 compared with theoretical luminosity functions wherein the effect 
of the IMF and the dependence on mass of the main sequence lifetime of a
star are included.

{\sc Fig}.\ 18.--- $M_V$ versus $(V-I)$ for stars in all of the annuli that
have photometric errors $< 0.5$ mags. in each color and $\chi < 2$. An extended
DA WD cooling sequence for C-core composition is included in the right panel to indicate the ages of
the cooling WDs. WDs as old as 9 Gyr appear to have been detected in M4.   
  
\begin{planotable}{llrrllrrrrrr}
\tablewidth{0pt}
\tablecaption{M4 White Dwarfs} \label{tbl-1}
\small
\tablehead
{\colhead{F} &
\colhead{C} &
\colhead{X} &
\colhead{Y} &
\colhead{RA} &
\colhead{DEC} &
\colhead{U} &
\colhead{{$\sigma$(U)}} &
\colhead{V} &
\colhead{{$\sigma$(V)}} &
\colhead{I} &
\colhead{{$\sigma$(I)}}} 
\startdata 
0 &1    &496.143    &344.045  &245.90711  &--26.52547     &25.807     &1.344
 &24.084     &0.053     &23.576     &0.057\nl
0 &1    &362.712    &469.314  &245.90763  &--26.52320     &26.342     &1.049
 &22.112     &0.035     &21.873     &0.037\nl
0 &1    &603.582    &677.165  &245.90318  &--26.52282     &26.322     &0.767
 &24.255     &0.047     &23.930     &0.076\nl
0 &1    &380.025    & 67.895  &245.91069  &--26.52746     &25.294     &0.391
 &24.457     &0.043     &24.028     &0.140\nl
0 &1    &737.100    &133.400  &245.90606  &--26.52937     &26.277     &0.059
 &26.086     &0.173     &24.993     &0.177\nl
0 &1    &351.673    &188.520  &245.91004  &--26.52602     &26.155     &0.381
 &25.256     &0.093     &24.559     &0.172\nl
0 &1    &494.198    &227.122  &245.90809  &--26.52666     &24.676     &9.999
 &24.907     &0.052     &24.167     &0.032\nl
0 &1    &531.399    &230.733  &245.90762  &--26.52689     &25.774     &0.941
 &24.449     &0.064     &23.688     &0.201\nl
0 &1    &368.113    &292.171  &245.90901  &--26.52507     &26.385     &9.999
 &25.496     &0.100     &24.532     &0.081\nl
0 &1    & 49.368    &304.191  &245.91256  &--26.52263     &25.821     &0.779
 &25.395     &0.116     &24.532     &0.102\nl
0 &1    &192.136    &436.617  &245.90986  &--26.52230     &25.666     &0.596
 &24.130     &0.054     &23.691     &0.062\nl
0 &1    &778.726    &623.622  &245.90161  &--26.52465     &26.945     &0.374
 &26.553     &0.291     &25.326     &0.396\nl
0 &1    &280.881    &643.384  &245.90716  &--26.52082     &25.765     &1.268
 &25.189     &0.094     &24.351     &0.109\nl
0 &1    &578.181    &520.223  &245.90473  &--26.52425     &25.725     &2.473
 &24.689     &0.080     &24.071     &0.143\nl
0 &2    &475.710    &108.076  &245.90911  &--26.51432     &23.550     &0.072
 &23.734     &0.107     &23.427     &0.099\nl
0 &2    &322.573    &187.614  &245.91384  &--26.51647     &24.117     &0.235
 &23.827     &0.090     &23.442     &0.100\nl
0 &2    &670.463    &265.761  &245.90956  &--26.50742     &23.004     &0.152
 &22.899     &0.074     &22.401     &0.060\nl
0 &2    &764.570    &388.489  &245.91093  &--26.50335     &24.304     &0.201
 &23.919     &0.107     &23.399     &0.077\nl
0 &2    &184.601    & 33.322  &245.91249  &--26.52201     &24.253     &0.232
 &24.052     &0.131     &23.531     &0.174\nl
0 &2    &430.132    &173.199  &245.91156  &--26.51429     &26.418     &1.594
 &26.124     &0.230     &24.906     &0.270\nl
0 &2    &548.990    &306.412  &245.91275  &--26.50947     &26.155     &1.868
 &26.261     &0.414     &24.820     &0.258\nl
0 &2    &767.149    &326.185  &245.90934  &--26.50429     &26.497     &0.580
 &26.107     &0.263     &25.015     &0.410\nl
0 &2    &621.544    &336.663  &245.91220  &--26.50736     &26.898     &0.394
 &26.792     &0.442     &25.878     &0.329\nl
0 &2    &616.555    &341.072  &245.91239  &--26.50740     &25.385     &0.426
 &24.956     &0.127     &24.150     &0.149\nl
0 &2    &655.539    &379.733  &245.91267  &--26.50591     &25.706     &0.479
 &25.442     &0.106     &24.701     &0.088\nl
0 &2    &537.197    &381.248  &245.91484  &--26.50853     &26.597     &0.853
 &26.681     &0.294     &25.253     &0.202\nl
0 &2    &685.570    &456.725  &245.91406  &--26.50400     &25.769     &0.613
 &25.675     &0.187     &24.799     &0.186\nl
0 &2    &744.612    &524.205  &245.91468  &--26.50161     &25.867     &0.519
 &26.090     &0.315     &25.074     &0.210\nl
0 &2    &679.611    &525.742  &245.91589  &--26.50302     &25.670     &0.540
 &25.233     &0.138     &24.361     &0.122\nl
0 &2    &784.167    &542.302  &245.91442  &--26.50045     &24.809     &0.249
 &24.665     &0.135     &24.160     &0.156\nl
0 &3    &412.694    &713.550  &245.93618  &--26.53406     &23.112     &0.117
 &23.166     &0.065     &23.007     &0.067\nl
0 &3    &596.394    &779.435  &245.94194  &--26.53258     &21.885     &0.066
 &22.415     &0.040     &22.161     &0.023\nl
0 &3    &506.072    &717.885  &245.93859  &--26.53266     &25.028     &0.280
 &24.490     &0.080     &23.703     &0.112\nl
0 &3    &479.387    &782.073  &245.93907  &--26.53451     &25.904     &0.762
 &26.154     &0.188     &25.156     &0.190\nl
0 &4    &131.966    &322.229  &245.90930  &--26.53149     &24.240     &0.183
 &23.926     &0.092     &23.281     &0.038\nl
0 &4    &550.987    &339.503  &245.91625  &--26.54125     &23.776     &0.208
 &23.735     &0.046     &23.392     &0.048\nl
0 &4    &109.239    &359.538  &245.90795  &--26.53157     &23.758     &0.105
 &23.711     &0.094     &23.417     &0.138\nl
0 &4    &310.042    &375.391  &245.91108  &--26.53636     &21.630     &0.056
 &22.075     &0.063     &21.938     &0.043\nl
0 &4    &258.894    &473.253  &245.90771  &--26.53675     &23.045     &0.148
 &23.340     &0.101     &23.108     &0.084\nl
0 &4    &431.768    &475.534  &245.91070  &--26.54071     &23.356     &0.163
 &23.411     &0.073     &23.154     &0.093\nl
0 &4    &410.933    &740.087  &245.90365  &--26.54440     &24.268     &0.125
 &24.102     &0.068     &23.653     &0.135\nl
0 &4    &712.836    &235.416  &245.92172  &--26.54326     &26.817     &9.999
 &27.181     &0.442     &25.796     &0.226\nl
0 &4    &771.054    &315.314  &245.92073  &--26.54583     &25.963     &0.346
 &26.547     &0.305     &25.309     &0.467\nl
0 &4    &721.709    &349.379  &245.91900  &--26.54526     &26.397     &0.752
 &26.578     &0.239     &25.091     &0.322\nl
0 &4    &647.282    &421.195  &245.91587  &--26.54472     &27.272     &0.903
 &26.618     &0.498     &25.064     &0.400\nl
0 &4    &576.599    &471.804  &245.91336  &--26.54393     &24.942     &0.413
 &24.267     &0.107     &23.535     &0.094\nl
0 &4    &635.669    &472.746  &245.91438  &--26.54527     &25.967     &0.547
 &25.619     &0.221     &24.793     &0.243\nl
0 &4    &589.519    &496.091  &245.91297  &--26.54460     &26.534     &0.323
 &26.373     &0.226     &25.132     &0.250\nl
0 &4    &784.239    &498.760  &245.91632  &--26.54901     &26.384     &0.771
 &26.065     &0.240     &25.034     &0.364\nl
0 &4    &635.871    &499.790  &245.91369  &--26.54570     &25.769     &0.494
 &25.495     &0.117     &24.597     &0.112\nl
0 &4    &632.871    &593.176  &245.91128  &--26.54710     &27.376     &4.721
 &26.917     &0.452     &24.983     &0.220\nl
0 &4    &643.786    &600.402  &245.91128  &--26.54746     &26.015     &1.169
 &25.629     &0.218     &24.792     &0.128\nl
0 &4    &686.579    &782.843  &245.90745  &--26.55125     &24.276     &0.225
 &24.029     &0.062     &23.618     &0.093\nl
0 &4    &289.603    &784.965  &245.90040  &--26.54236     &22.583     &0.076
 &22.935     &0.060     &22.687     &0.062\nl
0 &4    &432.300    &516.846  &245.90966  &--26.54137     &25.587     &0.502
 &25.027     &0.120     &24.149     &0.316\nl
0 &4    &312.866    &587.359  &245.90577  &--26.53978     &25.580     &0.737
 &25.250     &0.158     &24.528     &0.262\nl
1 &1    &349.394    &277.235  &245.92100  &--26.51353     &25.723     &0.475
 &23.235     &0.044     &22.866     &0.049\nl
1 &1    &430.304    &291.978  &245.91996  &--26.51397     &25.613     &2.206
 &24.177     &0.070     &23.654     &0.060\nl
1 &1    &586.114    &453.700  &245.91687  &--26.51344     &26.501     &0.476
 &23.824     &0.059     &23.414     &0.069\nl
1 &1    &522.663    &486.096  &245.91732  &--26.51265     &25.290     &0.273
 &23.783     &0.051     &23.316     &0.040\nl
1 &1    &135.349    &324.071  &245.92308  &--26.51150     &26.413     &0.124
 &24.031     &0.051     &23.611     &0.070\nl
1 &1    &475.811    & 78.225  &245.92117  &--26.51648     &25.709     &1.143
 &26.467     &0.164     &25.382     &0.301\nl
1 &1    &602.277    &114.687  &245.91943  &--26.51702     &25.362     &1.257
 &26.027     &0.134     &24.800     &0.134\nl
1 &1    &673.158    &205.174  &245.91789  &--26.51661     &26.552     &0.453
 &25.263     &0.059     &24.439     &0.076\nl
1 &1    &292.554    &214.222  &245.92216  &--26.51377     &26.225     &0.697
 &26.081     &0.146     &25.045     &0.187\nl
1 &1    &615.985    &328.506  &245.91754  &--26.51494     &26.762     &0.049
 &25.998     &0.104     &25.084     &0.149\nl
1 &1    &137.536    &355.195  &245.92280  &--26.51120     &25.660     &1.500
 &24.766     &0.060     &24.111     &0.041\nl
1 &1    &593.278    &364.668  &245.91751  &--26.51441     &26.982     &0.533
 &27.741     &0.494     &26.018     &0.452\nl
1 &1    &360.138    &375.973  &245.92008  &--26.51260     &25.629     &0.849
 &25.194     &0.082     &24.466     &0.106\nl
1 &1    &564.521    &389.121  &245.91763  &--26.51395     &27.683     &9.902
 &27.454     &0.350     &25.640     &0.300\nl
1 &1    &491.151    &403.723  &245.91835  &--26.51327     &26.275     &1.458
 &23.930     &0.053     &23.410     &0.044\nl
1 &1    &126.262    &604.305  &245.92091  &--26.50857     &25.667     &0.597
 &23.962     &0.099     &23.608     &0.120\nl
1 &1    &438.086    &634.244  &245.91709  &--26.51052     &26.880     &0.931
 &27.631     &0.341     &25.989     &0.414\nl
1 &1    &695.747    &750.784  &245.91323  &--26.51120     &26.771     &2.659
 &26.591     &0.166     &25.034     &0.222\nl
1 &1    &156.746    &381.055  &245.92237  &--26.51107     &26.489     &0.732
 &27.275     &0.244     &25.944     &0.394\nl
1 &1    &260.636    &481.355  &245.92036  &--26.51079     &25.197     &0.551
 &24.711     &0.080     &24.017     &0.085\nl
1 &1    &748.344    &639.979  &245.91352  &--26.51271     &25.807     &1.088
 &26.499     &0.194     &25.417     &0.325\nl
1 &2    &155.110    &172.465  &245.92811  &--26.50892     &23.150     &0.079
 &23.342     &0.056     &23.010     &0.055\nl
1 &2    &594.307    &231.544  &245.92170  &--26.49819     &22.224     &0.068
 &22.641     &0.039     &22.354     &0.049\nl
1 &2    & 57.763    &306.117  &245.93318  &--26.50894     &24.109     &0.190
 &24.034     &0.126     &23.462     &0.063\nl
1 &2    &373.536    &347.314  &245.92854  &--26.50125     &24.524     &0.407
 &24.271     &0.069     &23.595     &0.140\nl
1 &2    & 80.782    &412.613  &245.93542  &--26.50672     &24.278     &0.117
 &24.324     &0.072     &23.710     &0.095\nl
1 &2    &111.330    &598.374  &245.93949  &--26.50305     &24.293     &0.158
 &24.213     &0.075     &23.479     &0.083\nl
1 &2    &286.338    &603.237  &245.93649  &--26.49908     &22.315     &0.096
 &22.625     &0.045     &22.267     &0.029\nl
1 &2    &324.737    &629.303  &245.93644  &--26.49781     &24.232     &0.176
 &24.007     &0.054     &23.515     &0.056\nl
1 &2    &162.090    &701.645  &245.94113  &--26.50027     &22.840     &0.145
 &23.111     &0.110     &22.695     &0.119\nl
1 &2    &788.443    &794.648  &245.93220  &--26.48495     &23.827     &0.347
 &23.598     &0.049     &23.168     &0.082\nl
1 &2    &593.449    & 90.630  &245.91822  &--26.50047     &23.683     &0.155
 &23.490     &0.081     &22.930     &0.065\nl
1 &2    &361.919    & 78.734  &245.92207  &--26.50582     &25.396     &0.460
 &25.639     &0.218     &24.723     &0.349\nl
1 &2    &340.592    &103.637  &245.92308  &--26.50590     &25.498     &0.521
 &25.494     &0.143     &24.548     &0.115\nl
1 &2    &714.394    &137.017  &245.91723  &--26.49704     &26.608     &0.557
 &26.205     &0.165     &25.101     &0.136\nl
1 &2    &746.224    &151.223  &245.91702  &--26.49611     &23.867     &0.222
 &23.957     &0.080     &23.466     &0.117\nl
1 &2    &216.170    &172.611  &245.92701  &--26.50756     &25.636     &0.548
 &26.163     &0.192     &24.958     &0.148\nl
1 &2    & 88.418    &195.916  &245.92990  &--26.51002     &25.039     &0.322
 &24.451     &0.114     &23.968     &0.139\nl
1 &2    &115.700    &226.671  &245.93017  &--26.50893     &25.890     &0.673
 &25.630     &0.143     &24.866     &0.330\nl
1 &2    &600.651    &253.046  &245.92211  &--26.49770     &25.658     &0.453
 &25.415     &0.108     &24.705     &0.177\nl
1 &2    & 97.713    &283.199  &245.93190  &--26.50842     &25.590     &0.308
 &25.385     &0.137     &24.545     &0.205\nl
1 &2    &472.723    &471.692  &245.92987  &--26.49704     &25.390     &0.289
 &24.992     &0.114     &24.202     &0.199\nl
1 &2    &156.768    &483.856  &245.93583  &--26.50389     &26.550     &0.894
 &26.027     &0.273     &24.656     &0.205\nl
1 &2    &587.031    &488.731  &245.92823  &--26.49422     &26.150     &0.456
 &26.136     &0.322     &24.889     &0.200\nl
1 &2    &625.878    &496.426  &245.92773  &--26.49323     &26.184     &0.911
 &26.446     &0.304     &25.476     &0.327\nl
1 &2    &349.245    &511.384  &245.93307  &--26.49916     &26.237     &0.755
 &25.778     &0.178     &25.080     &0.296\nl
1 &2    &147.867    &515.218  &245.93678  &--26.50358     &24.803     &0.125
 &24.669     &0.102     &24.312     &0.109\nl
1 &2    &756.205    &631.259  &245.92875  &--26.48822     &26.967     &0.533
 &27.674     &0.450     &26.265     &0.444\nl
1 &2    &294.998    &665.264  &245.93786  &--26.49789     &27.245     &2.730
 &28.364     &0.413     &26.042     &0.495\nl
1 &2    &237.385    &725.273  &245.94038  &--26.49821     &24.006     &0.186
 &23.940     &0.063     &23.441     &0.077\nl
1 &2    &236.032    &790.145  &245.94200  &--26.49720     &26.302     &1.116
 &26.099     &0.231     &24.887     &0.233\nl
1 &2    &334.299    &108.831  &245.92330  &--26.50596     &26.010     &0.643
 &25.950     &0.206     &25.153     &0.468\nl
1 &2    &354.561    &138.707  &245.92369  &--26.50503     &25.350     &0.362
 &24.970     &0.147     &24.212     &0.137\nl
1 &3    &439.010    & 92.945  &245.93733  &--26.50819     &23.923     &0.128
 &23.873     &0.040     &23.468     &0.059\nl
1 &3    &610.590    &190.707  &245.94335  &--26.50764     &23.541     &0.083
 &23.557     &0.038     &23.175     &0.067\nl
1 &3    &594.596    &224.457  &245.94355  &--26.50864     &22.476     &0.044
 &22.864     &0.029     &22.472     &0.074\nl
1 &3    &258.688    &398.806  &245.93826  &--26.51789     &23.164     &0.104
 &23.356     &0.056     &23.055     &0.071\nl
1 &3    &745.151    &490.305  &245.95202  &--26.51219     &23.021     &0.075
 &23.505     &0.043     &23.337     &0.044\nl
1 &3    &687.013    &543.497  &245.95153  &--26.51430     &22.828     &0.033
 &23.089     &0.035     &22.819     &0.049\nl
1 &3    & 69.732    &550.776  &245.93626  &--26.52428     &25.041     &0.344
 &24.772     &0.058     &24.062     &0.089\nl
1 &3    &332.595    &560.433  &245.94299  &--26.52033     &24.600     &0.276
 &24.163     &0.054     &23.671     &0.097\nl
1 &3    &192.072    &567.610  &245.93960  &--26.52272     &24.716     &0.317
 &24.379     &0.060     &23.767     &0.078\nl
1 &3    &252.974    &655.715  &245.94270  &--26.52372     &23.281     &0.116
 &23.435     &0.035     &23.091     &0.078\nl
1 &3    &732.149    &718.742  &245.95573  &--26.51749     &23.244     &0.125
 &23.350     &0.042     &23.026     &0.059\nl
1 &3    &231.610    & 63.617  &245.93164  &--26.51084     &23.217     &0.152
 &23.297     &0.053     &22.859     &0.042\nl
1 &3    &382.830    &108.107  &245.93620  &--26.50942     &26.257     &0.607
 &26.045     &0.131     &25.083     &0.177\nl
1 &3    &139.245    &133.553  &245.93058  &--26.51387     &26.527     &0.687
 &25.929     &0.165     &24.946     &0.264\nl
1 &3    &503.679    &134.166  &245.93968  &--26.50807     &25.746     &0.437
 &25.469     &0.097     &24.621     &0.080\nl
1 &3    &638.194    &139.601  &245.94313  &--26.50607     &24.949     &0.138
 &24.762     &0.071     &24.208     &0.160\nl
1 &3    &368.116    &208.699  &245.93760  &--26.51189     &26.795     &0.196
 &26.326     &0.269     &25.493     &0.365\nl
1 &3    &365.608    &256.783  &245.93840  &--26.51301     &25.620     &0.564
 &25.248     &0.066     &24.392     &0.143\nl
1 &3    &749.597    &266.914  &245.94818  &--26.50714     &25.522     &0.528
 &25.218     &0.084     &24.510     &0.162\nl
1 &3    &660.734    &332.531  &245.94712  &--26.51000     &26.593     &0.447
 &27.114     &0.365     &25.698     &0.407\nl
1 &3    &244.283    &346.499  &245.93696  &--26.51695     &26.160     &0.532
 &26.993     &0.263     &25.135     &0.136\nl
1 &3    &432.803    &360.296  &245.94192  &--26.51425     &25.732     &0.592
 &25.183     &0.076     &24.452     &0.111\nl
1 &3    &224.606    &401.446  &245.93745  &--26.51850     &26.183     &0.516
 &26.922     &0.261     &25.795     &0.395\nl
1 &3    & 40.860    &558.464  &245.93569  &--26.52491     &26.835     &0.709
 &26.876     &0.280     &25.788     &0.225\nl
1 &3    &502.923    &641.896  &245.94870  &--26.51943     &26.616     &0.616
 &26.546     &0.226     &24.972     &0.216\nl
1 &3    &436.170    &672.737  &245.94758  &--26.52118     &25.866     &0.294
 &26.140     &0.185     &24.969     &0.224\nl
1 &3    &189.628    &701.455  &245.94194  &--26.52573     &26.929     &0.829
 &26.690     &0.239     &25.559     &0.366\nl
1 &3    &505.508    &713.059  &245.95002  &--26.52097     &26.210     &0.871
 &25.687     &0.151     &24.661     &0.087\nl
1 &3    & 49.299    &770.546  &245.93968  &--26.52944     &26.500     &0.901
 &26.461     &0.205     &25.420     &0.181\nl
1 &3    &387.762    &772.659  &245.94813  &--26.52417     &23.911     &0.096
 &23.867     &0.041     &23.470     &0.050\nl
1 &3    & 57.666    &565.498  &245.93623  &--26.52480     &25.037     &0.277
 &24.529     &0.059     &23.907     &0.072\nl
1 &3    &480.416    &632.276  &245.94796  &--26.51958     &24.964     &0.243
 &24.517     &0.080     &23.862     &0.063\nl
1 &4    &175.795    & 75.643  &245.92793  &--26.51705     &23.453     &0.159
 &23.479     &0.033     &23.010     &0.072\nl
1 &4    &786.603    & 89.207  &245.93823  &--26.53098     &23.386     &0.164
 &23.467     &0.109     &23.123     &0.076\nl
1 &4    &475.114    &101.041  &245.93255  &--26.52417     &24.035     &0.383
 &24.343     &0.142     &23.868     &0.250\nl
1 &4    & 98.263    &353.537  &245.91958  &--26.51965     &22.816     &0.077
 &22.922     &0.092     &22.573     &0.123\nl
1 &4    &511.988    &366.376  &245.92652  &--26.52917     &22.726     &0.091
 &22.824     &0.048     &22.619     &0.050\nl
1 &4    & 69.449    &406.796  &245.91773  &--26.51984     &23.753     &0.129
 &23.922     &0.090     &23.616     &0.134\nl
1 &4    &149.575    &406.909  &245.91914  &--26.52164     &22.199     &0.110
 &22.672     &0.048     &22.521     &0.098\nl
1 &4    &551.871    &468.205  &245.92466  &--26.53167     &24.000     &0.178
 &23.789     &0.104     &23.253     &0.063\nl
1 &4    & 53.621    &481.753  &245.91558  &--26.52067     &22.703     &0.143
 &23.148     &0.360     &22.786     &0.190\nl
1 &4    &426.446    &507.664  &245.92146  &--26.52947     &22.282     &0.032
 &22.482     &0.065     &22.302     &0.130\nl
1 &4    &582.420    &509.741  &245.92416  &--26.53301     &24.065     &0.189
 &24.091     &0.102     &23.668     &0.151\nl
1 &4    &485.875    & 56.592  &245.93385  &--26.52372     &26.562     &9.999
 &26.145     &0.219     &25.122     &0.210\nl
1 &4    &431.788    & 73.207  &245.93249  &--26.52276     &25.163     &0.578
 &25.188     &0.160     &24.462     &0.297\nl
1 &4    &127.921    &111.954  &245.92617  &--26.51655     &25.322     &0.565
 &25.133     &0.207     &24.216     &0.133\nl
1 &4    &425.910    &187.107  &245.92953  &--26.52441     &27.013     &0.450
 &27.106     &0.408     &25.913     &0.249\nl
1 &4    &252.792    &193.502  &245.92633  &--26.52061     &24.919     &0.356
 &24.606     &0.114     &24.147     &0.150\nl
1 &4    &597.301    &192.798  &245.93238  &--26.52836     &26.010     &1.918
 &25.930     &0.260     &24.868     &0.203\nl
1 &4    & 79.875    &202.238  &245.92306  &--26.51688     &24.626     &0.223
 &24.618     &0.230     &23.816     &0.182\nl
1 &4    &479.737    &202.983  &245.93008  &--26.52587     &26.082     &0.940
 &25.363     &0.227     &24.382     &0.128\nl
1 &4    &381.679    &244.865  &245.92730  &--26.52432     &26.655     &0.566
 &27.196     &0.400     &25.786     &0.350\nl
1 &4    &675.266    &245.385  &245.93243  &--26.53094     &24.788     &0.710
 &25.411     &0.418     &24.774     &0.258\nl
1 &4    &530.809    &277.742  &245.92909  &--26.52820     &25.697     &0.122
 &25.675     &0.125     &24.650     &0.132\nl
1 &4    &562.166    &557.753  &245.92259  &--26.53331     &25.701     &0.316
 &25.564     &0.153     &24.678     &0.157\nl
1 &4    & 55.253    &597.785  &245.91270  &--26.52253     &25.825     &0.259
 &25.657     &0.179     &24.663     &0.336\nl
1 &4    &641.403    &650.327  &245.92166  &--26.53652     &24.299     &0.279
 &24.029     &0.088     &23.545     &0.099\nl
1 &4    &733.349    &491.053  &245.92727  &--26.53610     &22.652     &0.099
 &22.971     &0.071     &22.688     &0.067\nl
6 &1    &109.020    &695.087  &245.97895  &--26.53500     &.....      &.....
 &28.649     &0.235     &26.935     &0.239\nl
6 &1    &617.842    &763.431  &245.97259  &--26.53798     &.....      &.....
 &26.587     &0.130     &25.246     &0.150\nl
6 &1    &647.847    &734.535  &245.97248  &--26.53849     &.....      &.....
 &27.841     &0.296     &26.118     &0.418\nl
6 &1    &792.862    &664.476  &245.97142  &--26.54026     &.....      &.....
 &24.265     &0.054     &23.875     &0.130\nl
6 &1    &786.000    &330.415  &245.97416  &--26.54362     &.....      &.....
 &27.162     &0.136     &25.869     &0.394\nl
6 &1    &119.092    &548.364  &245.98002  &--26.53656     &.....      &.....
 &27.588     &0.263     &26.132     &0.415\nl
6 &1    &151.445    &436.138  &245.98057  &--26.53794     &.....      &.....
 &27.686     &0.349     &26.615     &0.464\nl
6 &1    &400.669    &187.882  &245.97973  &--26.54229     &.....      &.....
 &28.201     &0.423     &25.858     &0.341\nl
6 &1    &446.760    &587.896  &245.97595  &--26.53853     &.....      &.....
 &28.963     &0.431     &26.778     &0.370\nl
6 &1    & 89.280    &119.655  &245.98382  &--26.54073     &.....      &.....
 &28.055     &0.486     &26.965     &0.433\nl
6 &2    &213.671    &440.346  &245.99232  &--26.53079     &.....      &.....
 &24.794     &0.055     &24.174     &0.056\nl
6 &2    &501.645    &226.109  &245.98181  &--26.52782     &.....      &.....
 &27.130     &0.215     &25.448     &0.220\nl
6 &2    &364.132    &350.366  &245.98738  &--26.52889     &.....      &.....
 &25.013     &0.071     &24.490     &0.100\nl
6 &2    &218.394    &353.329  &245.99006  &--26.53209     &.....      &.....
 &27.510     &0.303     &26.000     &0.308\nl
6 &2    &296.210    &424.047  &245.99043  &--26.52922     &.....      &.....
 &26.407     &0.164     &25.202     &0.303\nl
6 &2    &155.546    &432.165  &245.99316  &--26.53222     &.....      &.....
 &27.713     &0.272     &26.129     &0.417\nl
6 &2    &687.260    &450.714  &245.98408  &--26.52008     &.....      &.....
 &27.791     &0.316     &25.924     &0.417\nl
6 &2    &616.300    &163.544  &245.97821  &--26.52627     &.....      &.....
 &26.144     &0.136     &25.460     &0.302\nl
6 &2    &669.101    &487.989  &245.98533  &--26.51988     &.....      &.....
 &25.999     &0.118     &24.970     &0.256\nl
6 &2    &339.665    &656.623  &245.99543  &--26.52451     &.....      &.....
 &23.852     &0.054     &23.382     &0.033\nl
6 &2    &310.651    & 30.680  &245.98039  &--26.53520     &.....      &.....
 &25.046     &0.089     &24.455     &0.135\nl
6 &2    &743.250    &713.292  &245.98959  &--26.51468     &.....      &.....
 &25.020     &0.070     &24.317     &0.067\nl
6 &2    &688.943    &207.703  &245.97801  &--26.52394     &.....      &.....
 &23.563     &0.050     &23.284     &0.063\nl
6 &2    &584.775    &261.926  &245.98121  &--26.52539     &.....      &.....
 &27.762     &0.484     &26.105     &0.233\nl
6 &2    &346.208    &322.712  &245.98701  &--26.52973     &.....      &.....
 &25.254     &0.096     &24.414     &0.124\nl
6 &2    &115.407    &363.480  &245.99217  &--26.53421     &.....      &.....
 &24.599     &0.044     &24.007     &0.109\nl
6 &2    &372.111    &269.815  &245.98522  &--26.53001     &.....      &.....
 &27.146     &0.230     &25.552     &0.256\nl
6 &2    & 55.610    &392.199  &245.99395  &--26.53508     &.....      &.....
 &25.887     &0.121     &24.823     &0.116\nl
6 &2    & 80.709    &466.106  &245.99534  &--26.53333     &.....      &.....
 &24.491     &0.060     &23.826     &0.089\nl
6 &2    &428.026    &473.008  &245.98928  &--26.52549     &.....      &.....
 &27.819     &0.356     &26.035     &0.328\nl
6 &2    &581.605    &514.089  &245.98754  &--26.52141     &.....      &.....
 &24.985     &0.084     &24.003     &0.054\nl
6 &2    &443.064    &756.434  &245.99604  &--26.52062     &.....      &.....
 &26.491     &0.104     &25.396     &0.287\nl
6 &2    &545.114    &157.697  &245.97933  &--26.52795     &.....      &.....
 &26.066     &0.152     &24.969     &0.117\nl
6 &2    &381.549    &598.594  &245.99325  &--26.52450     &.....      &.....
 &25.399     &0.119     &24.375     &0.072\nl
6 &2    &766.326    &182.357  &245.97602  &--26.52264     &.....      &.....
 &24.189     &0.060     &23.698     &0.144\nl
6 &2    &187.376    &315.015  &245.98966  &--26.53340     &.....      &.....
 &26.668     &0.185     &25.688     &0.183\nl
6 &2    &771.348    &629.224  &245.98701  &--26.51539     &.....      &.....
 &27.322     &0.320     &26.049     &0.450\nl
6 &2    &645.273    &316.654  &245.98149  &--26.52316     &.....      &.....
 &25.616     &0.118     &24.984     &0.202\nl
6 &2    &150.198    &792.555  &246.00216  &--26.52654     &.....      &.....
 &26.335     &0.249     &24.931     &0.255\nl
6 &2    &770.776    &492.578  &245.98362  &--26.51757     &.....      &.....
 &27.718     &0.397     &25.552     &0.238\nl
6 &2    &772.211    &143.847  &245.97497  &--26.52313     &.....      &.....
 &27.706     &0.391     &26.191     &0.361\nl
6 &2    &147.354    & 37.054  &245.98349  &--26.53871     &.....      &.....
 &28.302     &0.444     &26.948     &0.252\nl
6 &2    &506.589    &606.140  &245.99119  &--26.52160     &.....      &.....
 &28.724     &0.386     &26.400     &0.277\nl
6 &3    &262.622    &593.039  &246.00041  &--26.54964     &.....      &.....
 &24.766     &0.075     &24.147     &0.111\nl
6 &3    &630.487    &217.970  &246.00292  &--26.53540     &.....      &.....
 &23.322     &0.034     &23.059     &0.054\nl
6 &3    &222.354    &333.591  &245.99478  &--26.54449     &.....      &.....
 &25.286     &0.066     &24.417     &0.097\nl
6 &3    &195.767    &340.510  &245.99423  &--26.54506     &.....      &.....
 &26.895     &0.208     &25.778     &0.328\nl
6 &3    &667.068    &417.219  &246.00738  &--26.53927     &.....      &.....
 &22.875     &0.028     &22.611     &0.058\nl
6 &3    &408.539    &463.328  &246.00175  &--26.54442     &.....      &.....
 &26.501     &0.171     &25.556     &0.324\nl
6 &3    &627.251    &541.452  &246.00860  &--26.54268     &.....      &.....
 &27.666     &0.482     &26.150     &0.374\nl
6 &3    &423.940    &550.251  &246.00369  &--26.54612     &.....      &.....
 &26.892     &0.158     &25.734     &0.190\nl
6 &3    &101.633    &581.690  &245.99620  &--26.55194     &.....      &.....
 &28.187     &0.379     &27.142     &0.022\nl
6 &3    &187.459    &189.255  &245.99134  &--26.54181     &.....      &.....
 &25.392     &0.087     &24.694     &0.068\nl
6 &3    &393.543    &257.209  &245.99769  &--26.54004     &.....      &.....
 &27.713     &0.171     &26.708     &0.451\nl
6 &3    &332.736    &348.005  &245.99780  &--26.54305     &.....      &.....
 &25.672     &0.112     &24.319     &0.081\nl
6 &3    &606.532    &225.956  &246.00247  &--26.53596     &.....      &.....
 &25.291     &0.090     &24.420     &0.128\nl
6 &3    &198.302    &611.158  &245.99913  &--26.55107     &.....      &.....
 &26.940     &0.198     &25.518     &0.193\nl
6 &3    &418.869    & 54.389  &245.99473  &--26.53512     &.....      &.....
 &26.916     &0.159     &25.636     &0.325\nl
6 &3    &744.329    & 60.791  &246.00296  &--26.53014     &.....      &.....
 &27.630     &0.461     &26.216     &0.478\nl
6 &3    &727.209    &605.294  &246.01220  &--26.54251     &.....      &.....
 &27.447     &0.359     &26.069     &0.293\nl
6 &3    &260.083    &756.515  &246.00326  &--26.55331     &.....      &.....
 &28.444     &0.471     &25.997     &0.381\nl
6 &3    &443.506    &256.110  &245.99893  &--26.53922     &.....      &.....
 &27.988     &0.239     &27.084     &0.431\nl
6 &3    &512.311    &785.385  &246.01005  &--26.54994     &.....      &.....
 &24.011     &0.046     &23.562     &0.084\nl
6 &3    &329.639    & 80.680  &245.99296  &--26.53713     &.....      &.....
 &26.773     &0.173     &25.254     &0.164\nl
6 &3    &372.806    &726.795  &246.00555  &--26.55086     &.....      &.....
 &28.247     &0.465     &26.754     &0.319\nl
6 &3    &333.206    &556.964  &246.00154  &--26.54772     &.....      &.....
 &28.243     &0.463     &27.226     &0.351\nl
6 &3    &525.824    & 61.948  &245.99754  &--26.53359     &.....      &.....
 &28.779     &0.398     &27.736     &0.040\nl
6 &4    &384.522    &396.847  &245.98210  &--26.55425     &.....      &.....
 &25.471     &0.080     &24.456     &0.094\nl
6 &4    &292.027    &470.782  &245.97862  &--26.55332     &.....      &.....
 &24.063     &0.048     &23.594     &0.091\nl
6 &4    & 74.221    &478.114  &245.97462  &--26.54854     &.....      &.....
 &24.763     &0.074     &24.132     &0.062\nl
6 &4    &267.537    &610.828  &245.97467  &--26.55498     &.....      &.....
 &25.634     &0.098     &24.820     &0.134\nl
6 &4    &375.318    &651.784  &245.97552  &--26.55805     &.....      &.....
 &24.606     &0.073     &24.003     &0.075\nl
6 &4    &318.687    &689.369  &245.97358  &--26.55736     &.....      &.....
 &27.140     &0.187     &25.708     &0.247\nl
6 &4    &516.217    &349.434  &245.98562  &--26.55647     &.....      &.....
 &26.465     &0.184     &25.199     &0.363\nl
6 &4    &433.489    &254.303  &245.98657  &--26.55311     &.....      &.....
 &27.964     &0.391     &26.115     &0.250\nl
6 &4    &227.322    &551.512  &245.97544  &--26.55314     &.....      &.....
 &24.085     &0.059     &23.564     &0.047\nl
6 &4    &474.078    &757.921  &245.97460  &--26.56192     &.....      &.....
 &25.987     &0.118     &25.229     &0.173\nl
6 &4    &503.497    &709.015  &245.97634  &--26.56182     &.....      &.....
 &26.555     &0.153     &25.545     &0.192\nl
6 &4    &321.260    &795.249  &245.97100  &--26.55907     &.....      &.....
 &23.881     &0.049     &23.449     &0.101\nl
6 &4    &198.675    &547.455  &245.97505  &--26.55243     &.....      &.....
 &27.103     &0.274     &25.268     &0.173\nl
6 &4    & 54.743    &669.624  &245.96949  &--26.55112     &.....      &.....
 &25.083     &0.087     &24.219     &0.149\nl
6 &4    &737.255    &474.875  &245.98632  &--26.56340     &.....      &.....
 &24.188     &0.045     &23.629     &0.048\nl
6 &4    & 75.411    &141.795  &245.98309  &--26.54331     &.....      &.....
 &27.613     &0.315     &25.508     &0.399\nl
6 &4    &733.469    &284.282  &245.99105  &--26.56033     &.....      &.....
 &25.511     &0.128     &24.725     &0.175\nl
6 &4    &573.187    &739.180  &245.97682  &--26.56385     &.....      &.....
 &25.428     &0.092     &24.507     &0.214\nl
6 &4    & 50.351    &191.311  &245.98141  &--26.54352     &.....      &.....
 &24.129     &0.058     &23.698     &0.095\nl
6 &4    & 80.333    &140.315  &245.98321  &--26.54340     &.....      &.....
 &27.351     &0.274     &25.979     &0.386\nl
6 &4    &184.874    & 64.031  &245.98696  &--26.54455     &.....      &.....
 &26.729     &0.134     &25.706     &0.386\nl
6 &4    &715.109    &793.780  &245.97795  &--26.56784     &.....      &.....
 &25.770     &0.115     &24.810     &0.147\nl
6 &4    &120.982    &766.116  &245.96824  &--26.55412     &.....      &.....
 &28.571     &0.172     &27.472     &0.375\nl
6 &4    &256.210    &277.789  &245.98284  &--26.54948     &.....      &.....
 &27.950     &0.417     &26.691     &0.434\nl

\enddata 
\end{planotable}


\end{document}